\documentstyle[sprocl,psfig]{article}

\def\be{\begin{equation}}
\def\ee{\end{equation}}
\def\bea{\begin{eqnarray}}
\def\eea{\end{eqnarray}}

\begin{document}

\title{CONTINUUM QCD AND LIGHT MESONS}

\author{P. MARIS}

\address{Dept. of Physics, Kent State University,
Kent, OH 44242\\
E-mail: maris@faisun.kent.edu} 


\maketitle\abstracts{
The ladder-rainbow truncation of the set of Dyson--Schwinger equations
is used to study light mesons.  The parameters in the effective
interaction are constrained by the chiral condensate and $f_\pi$; the
current quark masses are fitted to $m_\pi$ and $m_K$.  The dressed
quark propagators are in qualitative agreement with recent lattice-QCD
results at low $q^2$ while having the correct perturbative behavior at
large $q^2$.  The resulting vector meson masses are within 5\% of the
experimental values.  The obtained electromagnetic form factors and
strong and electroweak coupling constants are also in good agreement
with the data.  At finite temperature, this truncation leads to a
mean-field chiral phase transition.  The spatial pion mass is almost
constant below this transition, but rises with $T$ close to and above
$T_c$.  The mass of its chiral partner, an idealized $\sigma$ meson,
decreases with $T$ until $T_c$, where it becomes degenerate with the
pion.}

\section{Introduction}
Our aim is to obtain the hadron mass spectrum and observables such as
coupling constants and form factors from the underlying theory, QCD.
The set of Dyson--Schwinger equations [DSEs] form a useful tool for
this purpose~\cite{Roberts:2000aa,Alkofer:2000wg}.  In rainbow-ladder
truncation, they have been successfully applied to calculate the
masses and decay constants of light pseudoscalar and vector
mesons~\cite{Maris:1997tm,Maris:1999nt}.  The dressed-quark
propagator, as obtained from its DSE, together with the
Bethe--Salpeter amplitude [BSA] and the $q\bar q\gamma$ vertex as
obtained from the homogeneous and inhomogeneous Bethe--Salpeter
equations [BSE] respectively, form the necessary elements for form
factor calculations in impulse
approximation~\cite{Maris:2000bh,Maris:2000sk}.  Extensions to other
mesons, and calculations of strong and electroweak decays and
branching ratios are straightforward~\cite{extensions,Ivanov:1999ms}.

At finite temperature chiral symmetry is expected to be restored, and
quarks (and gluons) become deconfined.  Since the properties of the
pion are closely tied to the dynamical breaking of chiral symmetry, an
elucidation of the $T$-dependence of these properties is important;
particularly since a prodigious number of pions is produced in heavy
ion collisions.  Encouraged by the success of the DSE approach to
zero-temperature pion physics, see in Sec.~\ref{modelcalc}, we have
extended this approach to nonzero temperature~\cite{Maris:2000ig},
and this is discussed in Sec.~\ref{finiteT}.

The DSE for the renormalized quark propagator in Euclidean space is
\begin{equation}
\label{gendse}
 S(p)^{-1} = i \, Z_2\, /\!\!\!p + Z_4\,m(\mu) + 
        Z_1 \int^\Lambda\!\!\!\frac{d^4q}{(2\pi)^4} \,g^2 D_{\mu\nu}(k) 
        \frac{\lambda^a}{2}\gamma_\mu S(q)\Gamma^a_\nu(q,p) \,,
\end{equation}
where $D_{\mu\nu}(k)$ is the dressed-gluon propagator,
$\Gamma^a_\nu(q;p)$ the dressed-quark-gluon vertex, and $k=p-q$.  
The most general solution of Eq.~(\ref{gendse}) has the form
\mbox{$S(p)^{-1} = i /\!\!\! p A(p^2) + B(p^2)$} and is renormalized 
at spacelike $\mu^2$ according to \mbox{$A(\mu^2)=1$} and
\mbox{$B(\mu^2)=m(\mu)$} with $m(\mu)$ the current quark mass.
The notation $\int^{\Lambda}$ represents a translationally invariant
regularization of the integral with $\Lambda$ the regularization
mass-scale; at the end of all calculations this regularization scale can
be removed.

Mesons are described by solutions of the homogeneous BSE
\begin{equation}
 \Gamma_H(p_+,p_-) = \int^\Lambda\!\!\!\frac{d^4q}{(2\pi)^4} \, 
        K(p,q;Q) \; S(q_+) \, \Gamma_H(q_+,q_-) \, S(q_-)\, ,
\label{homBSE}
\end{equation}
where $p_+ = p + \eta Q$ and $p_- = p - (1-\eta) Q$ are the outgoing
and incoming quark momenta respectively, and similarly for $q_\pm$.
The kernel $K$ is the renormalized, amputated $q\bar q$ scattering
kernel that is irreducible with respect to a pair of $q\bar q$ lines.
This equation has solutions at discrete values of $Q^2 = -m_H^2$,
where $m_H$ is the meson mass.  Together with the canonical
normalization condition for $q\bar q$ bound states, it completely
determines $\Gamma_H$, the bound state BSA.  The different types of
mesons, such as (pseudo-)scalar, (axial-)vector, and tensor mesons,
are characterized by different Dirac structures.  The most general
decomposition for pseudoscalar bound states is~\cite{Maris:1997tm}
\begin{eqnarray}
\label{genpion}
\Gamma_{PS}(k_+,k_-;P) &=& \gamma_5 \big[ i E(k^2;k\cdot P;\eta) + 
        \;/\!\!\!\! P \, F(k^2;k\cdot P;\eta) 
\nonumber \\
        && {} + \,/\!\!\!k \, G(k^2;k\cdot P;\eta) +
        \sigma_{\mu\nu}\,k_\mu P_\nu \,H(k^2;k\cdot P;\eta) \big]\,,
\end{eqnarray}
where the invariant amplitudes $E$, $F$, $G$ and $H$ are Lorentz
scalar functions of $k^2$ and $k\cdot P$.  Note that these functions
depend on the momentum partitioning parameter $\eta$; however,
physical observables are independent of this parameter.

The bound state BSA, together with the dressed quark propagators, 
are the necessary elements to calculate the electroweak decay
constants~\cite{Maris:1997tm,Maris:1999nt} and, in impulse
approximation, strong decays.  For describing other electroweak
processes one also needs the dressed $q\bar q\gamma$ and $q\bar q W$
vertices.  These vertices satisfy an inhomogeneous BSE: e.g. the
$q\bar q\gamma$ vertex \mbox{$\Gamma_\mu(p_+,p_-)$} satisfies
\begin{equation}
 \Gamma_\mu(p_+,p_-) = Z_2 \, \gamma_\mu + 
        \int^\Lambda\!\!\!\frac{d^4q}{(2\pi)^4} \, K(p,q;Q) 
        \;S(q_+) \, \Gamma_\mu(q_+,q_-) \, S(q_-)\, .
\label{verBSE}
\end{equation}
Solutions of the homogeneous version of Eq.~(\ref{verBSE}) define vector
meson bound states at timelike photon momenta \mbox{$Q^2=-m_{\rm v}^2$}.
It follows that $\Gamma_\mu(p_+,p_-)$ has poles at those locations.

\section{Model Calculations}
\label{modelcalc}
To solve the BSE, we use a ladder truncation, with an effective
quark-antiquark interaction that reduces to the perturbative running
coupling at large momenta~\cite{Maris:1997tm,Maris:1999nt}.  In
conjunction with the rainbow truncation for the quark DSE, the ladder
truncation preserves both the vector Ward--Takahashi identity [WTI] for
the $q\bar q\gamma$ vertex and the axial-vector WTI.  The latter ensures
the existence of massless pseudoscalar mesons connected with dynamical
chiral symmetry breaking~\cite{Maris:1997tm}.  In combination with
impulse approximation, this truncation satisfies electromagnetic current
conservation~\cite{Maris:2000sk}.

\subsection{Effective interaction}
The ladder truncation of the BSE, Eq.~(\ref{homBSE}), is
\begin{equation}
\label{ourBSEansatz}
        K(p,q;P) \to
        -{\cal G}(k^2)\, D_{\mu\nu}^{\rm free}(k)
        \textstyle{\frac{\lambda^i}{2}}\gamma_\mu \otimes
        \textstyle{\frac{\lambda^i}{2}}\gamma_\nu \,,
\end{equation}
where $D_{\mu\nu}^{\rm free}(k=p-q)$ is the free gluon propagator in
Landau gauge.  The corresponding rainbow truncation of the quark DSE,
Eq.~(\ref{gendse}), is given by \mbox{$\Gamma^i_\nu(q,p) \rightarrow 
\gamma_\nu\lambda^i/2$} together with \mbox{$g^2 D_{\mu \nu}(k) 
\rightarrow {\cal G}(k^2) D_{\mu\nu}^{\rm free}(k) $}.
This truncation was found to be particularly suitable for the flavor
octet pseudoscalar and vector mesons since the next-order
contributions in a quark-gluon skeleton graph expansion, have a
significant amount of cancellation between repulsive and attractive
corrections~\cite{Bender:1996bb}.  

The model is completely specified once a form is chosen for the
``effective coupling'' ${\cal G}(k^2)$.  We employ the
Ansatz~\cite{Maris:1999nt}
\begin{eqnarray}
\label{gvk2}
\frac{{\cal G}(k^2)}{k^2} &=&
        \frac{4\pi^2\, D \,k^2}{\omega^6} \, {\rm e}^{-k^2/\omega^2}
        + \frac{ 4\pi^2\, \gamma_m \; {\cal F}(k^2)}
        {\textstyle{\frac{1}{2}} \ln\left[\tau + 
        \left(1 + k^2/\Lambda_{\rm QCD}^2\right)^2\right]} \;,
\end{eqnarray}
with \mbox{$\gamma_m=12/(33-2N_f)$} and
\mbox{${\cal F}(s)=(1 - \exp\frac{-s}{4 m_t^2})/s$}.  
The ultraviolet behavior is chosen to be that of the QCD running
coupling $\alpha(k^2)$; the ladder-rainbow truncation then generates
the correct perturbative QCD structure of the DSE-BSE system of
equations.  The first term implements the strong infrared
enhancement
in the region \mbox{$0 < k^2 < 1\,{\rm GeV}^2$} phenomenologically
required~\cite{Hawes:1998cw} to produce a realistic
value~\cite{Leinweber:1997fn} for the chiral condensate of about
$(240\,{\rm GeV})^3$.  We use \mbox{$m_t=0.5\,{\rm GeV}$},
\mbox{$\tau={\rm e}^2-1$}, \mbox{$N_f=4$}, \mbox{$\Lambda_{\rm QCD} =
0.234\,{\rm GeV}$}, and a renormalization scale \mbox{$\mu=19\,{\rm
GeV}$} which is well into the perturbative
domain~\cite{Maris:1997tm,Maris:1999nt}.  The remaining parameters,
\mbox{$\omega = 0.4\,{\rm GeV}$} and
\mbox{$D=0.93\,{\rm GeV}^2$} along with the quark masses, are fitted to
give a good description of the chiral condensate, $m_{\pi/K}$ and
$f_{\pi}$, see Table~\ref{tab:ps}.  
\begin{table}[t]
\caption{Results for the light pseudoscalar mesons in GeV. \label{tab:ps}}
\vspace{0.2cm}
\begin{center}
\begin{tabular}{|l|ll|lll|lll|}
\hline
& \multicolumn{2}{|c|}{expt. data} 
& \multicolumn{3}{|c|}{our calculation} 
& \multicolumn{3}{|c|}{$\gamma_5$ amplitude $E$ only} \\ \hline
meson  &$m_H$&$f_H$&$m_H$&$f_H$&${\cal R}_H$&$m_H$&$f_H$&${\cal R}_H$ 
\\ \hline
chiral & 0.0   & --     & 0.0   & 0.090 & 1.00 & 0.0   & 0.067 & 1.52 \\
pion   & 0.139 & 0.0924 & 0.138 & 0.092 & 0.99 & 0.121 & 0.069 & 1.51 \\
kaon   & 0.497 & 0.112  & 0.496 & 0.109 & 0.99 & 0.436 & 0.081 & 1.43 \\
$s\bar s$& --  & --     & 0.696 & 0.129 & 0.99 & 0.589 & 0.099 & 1.44 
\\ \hline
\end{tabular}
\end{center}
\end{table}
Note that all invariant amplitudes of the BSA, see
Eq.~(\ref{genpion}), are needed to satisfy the axial-vector
WTI~\cite{Maris:1997tm}: ${\cal R}_H$ is the ratio of the residues at
the meson pole in the pseudovector and pseudoscalar vertices which
appear in this WTI, and should be equal to one.  From
Table~\ref{tab:ps} it is also clear that the pseudovector amplitudes
$F$ and $G$ do contribute to observables; furthermore, they dominate
the asymptotic behavior of the pion electromagnetic form
factor~\cite{Maris:1998hc}.

\subsection{Quark propagator}
Within this model, the quark propagator reduces to the perturbative
propagator in the ultraviolet region.  The mass function 
\mbox{$M(p^2) = B(p^2)/A(p^2)$} behaves for large $p^2$ like
\be
 M(p^2) \simeq \frac{\hat{m}_q}{\left(
        \frac{1}{2}\ln\left[\frac{p^2}{\Lambda_{\rm QCD}^2}
                \right]\right)^{\gamma_m}} \;,
\ee 
where $\hat{m}_q$ is the renormalization-point-independent
explicit chiral-symmetry-breaking mass.  In the chiral limit the
behavior is qualitatively different
\be
 M(p^2) \simeq \frac{2\pi^2\gamma_m}{3}\,
        \frac{-\,\langle \bar q q \rangle^0}{p^2 \left(
        \frac{1}{2}\ln\left[\frac{p^2}{\Lambda_{\rm QCD}^2}
                                        \right] \right)^{1-\gamma_m}}\,,
\ee
with $\langle \bar q q \rangle^0$ the renormalization-point-independent
chiral condensate~\cite{Maris:1997tm}.  At one-loop, it is related to
the renormalization-dependent condensate
\be
  \langle \bar q q \rangle^\mu = -Z_4 \, N_c\,\int^\Lambda\!\! 
        \frac{d^4p}{(2\pi)^4} {\rm Tr}[S_{\hbox{\scriptsize{chiral}}}(p)] \,,
\ee
via $\langle \bar q q \rangle^\mu = (\ln{\mu/\Lambda_{\rm
QCD}})^{\gamma_m} \, \langle \bar q q \rangle^0$.

\begin{figure}[t]
\psfig{figure=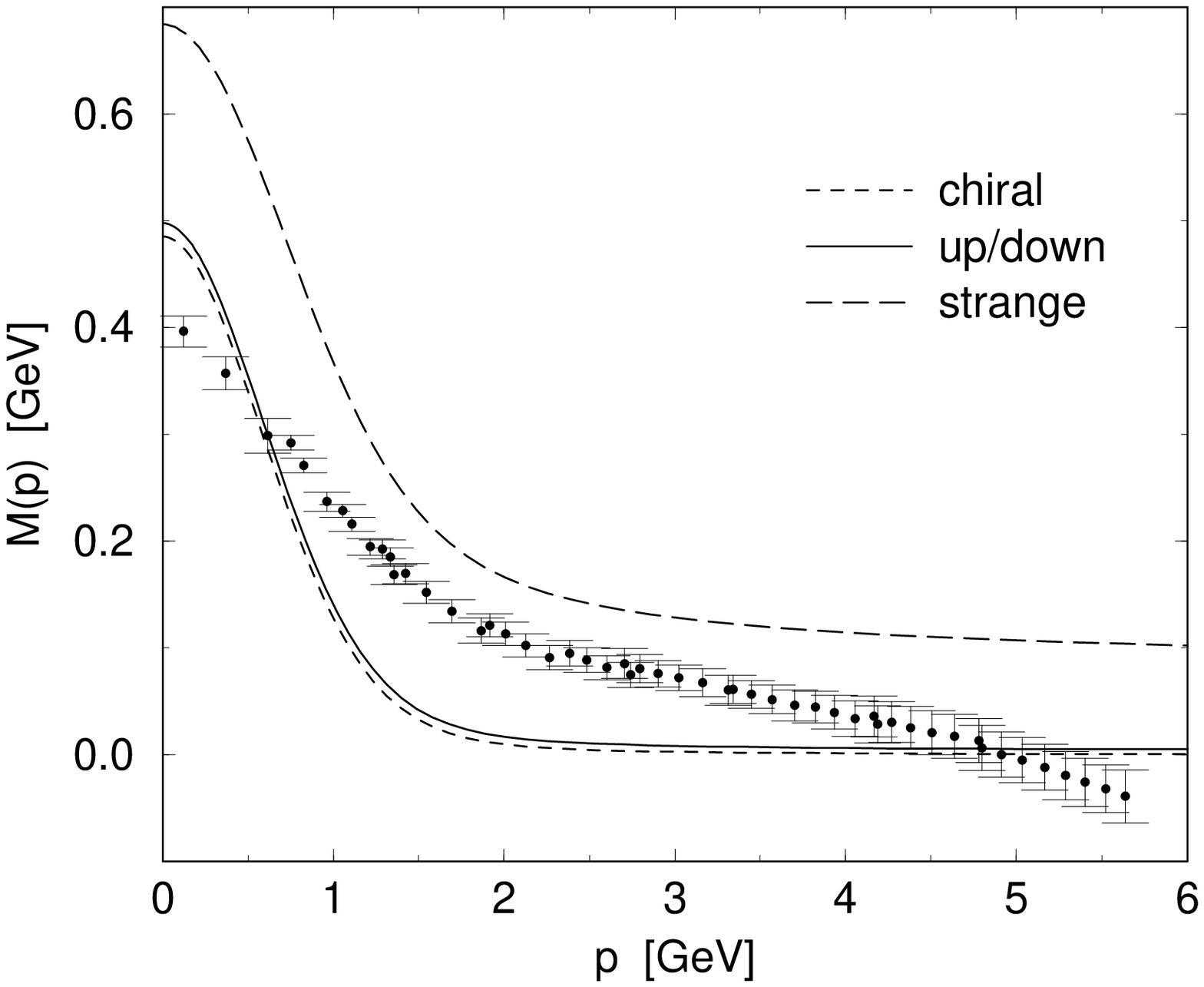,height=1.8in}
\vspace*{-1.8in}

\hspace*{2.4in}\psfig{figure=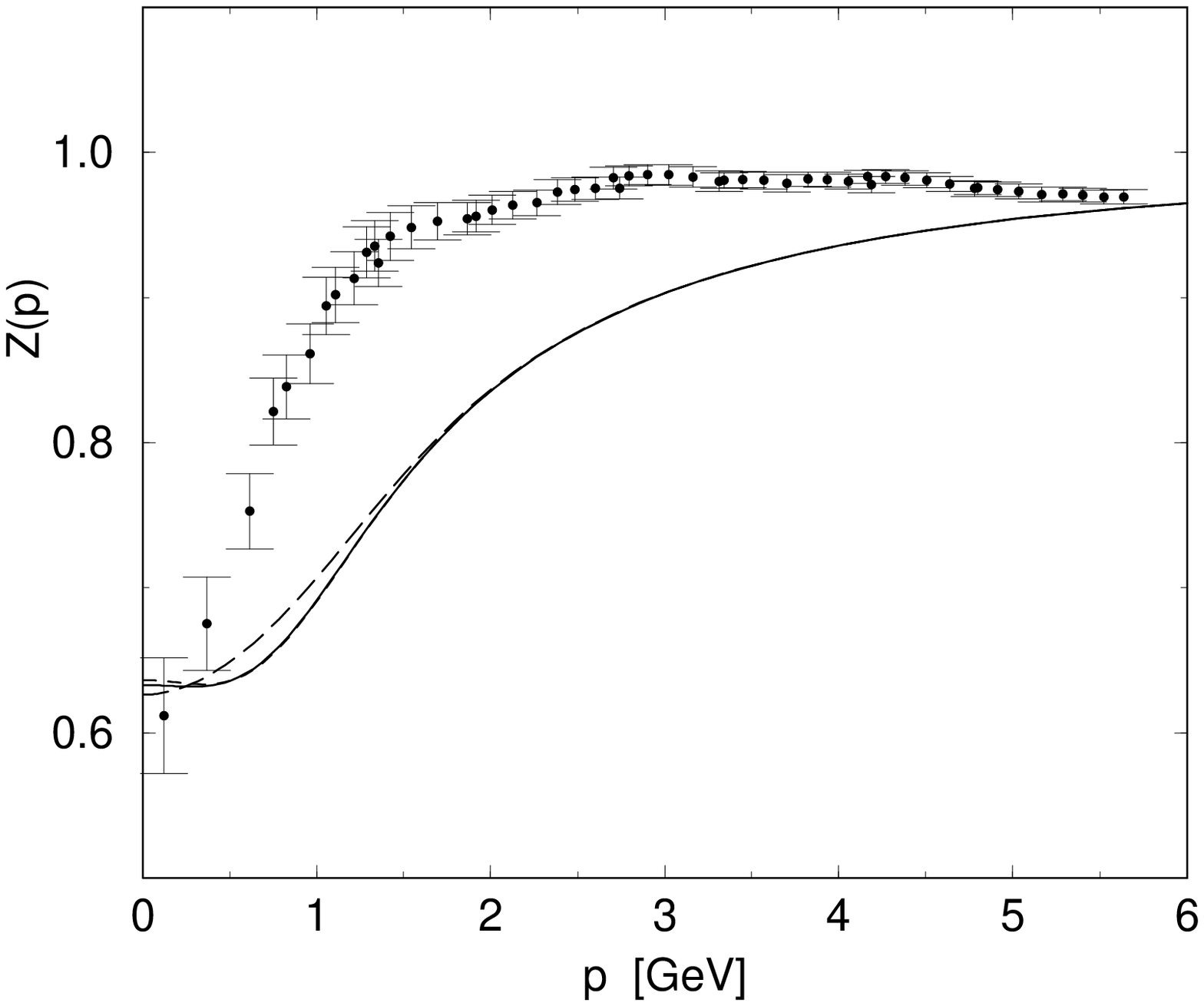,height=1.8in}
\caption{The quark functions $M(p)$ (left) and $Z(p)$ (right), compared 
to lattice data~\protect\cite{Skullerud:2000un}.  The lattice data
correspond to a bare quark mass of 112 MeV, which is between our
up/down and strange quark mass.  Note that the lattice data for $M(p)$
are unreliable above $p \approx 1.5$~GeV.
\label{fig:qrkprp}}
\end{figure}

In the infrared region both $Z(p^2)=1/A(p^2)$ and $M(p^2)$ deviate
significantly from the perturbative behavior, due to chiral symmetry
breaking.  Qualitatively, our results are similar to those obtained in
recent lattice simulations~\cite{Skullerud:2000un}, see
Fig.~\ref{fig:qrkprp}.
The strong enhancement of $M(p^2)$ in the infrared region is a
characteristic signal for dynamical chiral symmetry breaking.  Also
the behavior of $Z(p^2)$ is characteristic: at large $p^2$, $Z(p^2)=1$
up to logarithmic corrections, whereas DSE studies typically find
$Z(0)$ to be significantly smaller than one~\cite{Roberts:2000hi}.
This strongly decreased value of $Z(0)$ is now confirmed by lattice
data.  The main difference between our results and the lattice data is
the onset of the deviation from the perturbative behavior, both for
$M(p)$ and for $Z(p)$; this mass scale does depend on details of the
effective interaction ${\cal G}(k^2)$.

\subsection{Meson Masses and Decay Constants}
Within the same model, we can now solve the BSE for vector mesons, and
calculate the vector meson masses and decay constants.  The $\rho$ and
$\phi$ electroweak decay constants are related to the $\rho-\gamma$ and
$\phi-\gamma$ coupling constants respectively; the decay constant for
the $K^*$ can be extracted from $\tau$-decay into a neutrino and a $K^*$
via a $W$-boson.  The results of our model
calculations~\cite{Maris:1999nt} are shown in Table~\ref{tab:vectors},
\begin{table}[b]
\caption{Overview of results for light vector meson masses and 
decay constants, all in GeV. 
\label{tab:vectors}}
\vspace{0.2cm}
\begin{center}
\begin{tabular}{|l|llllll|} \hline
         &$m_\rho$&$f_\rho$&$m_{K^*}$&$f_{K^*}$&$m_\phi$&$f_\phi$ 
\\ \hline
expt. data~\protect\cite{PDG}      
                & 0.770 & 0.216 & 0.892 & 0.225 & 1.020 & 0.236 \\
calculation     & 0.742 & 0.207 & 0.936 & 0.241 & 1.072 & 0.259 
\\ \hline
\end{tabular}
\end{center}
\end{table}
and are in reasonable agreement with the data.  Corrections from
pseudoscalar meson loops are estimated to be small~\cite{rhomass},
and will generate a nonzero width for the vector mesons.

Although nature provides us with data for fixed quark masses only, it is
interesting to see how observables such as meson masses vary with
current quark mass, in particular for comparison with other
nonperturbative methods such as lattice QCD.
\begin{figure}[t]
\psfig{figure=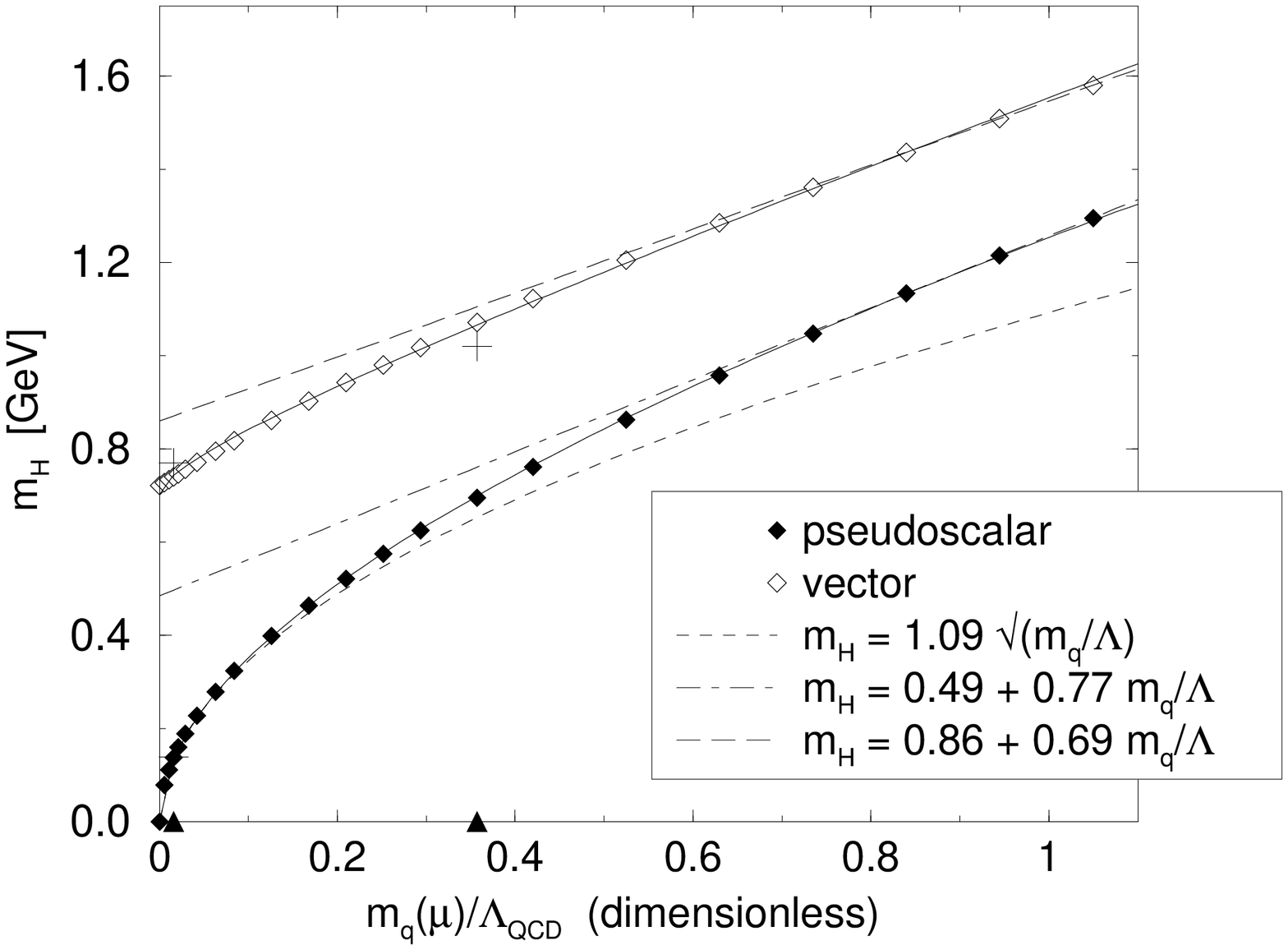,height=1.7in}
\vspace*{-1.7in}

\hspace*{2.4in}\psfig{figure=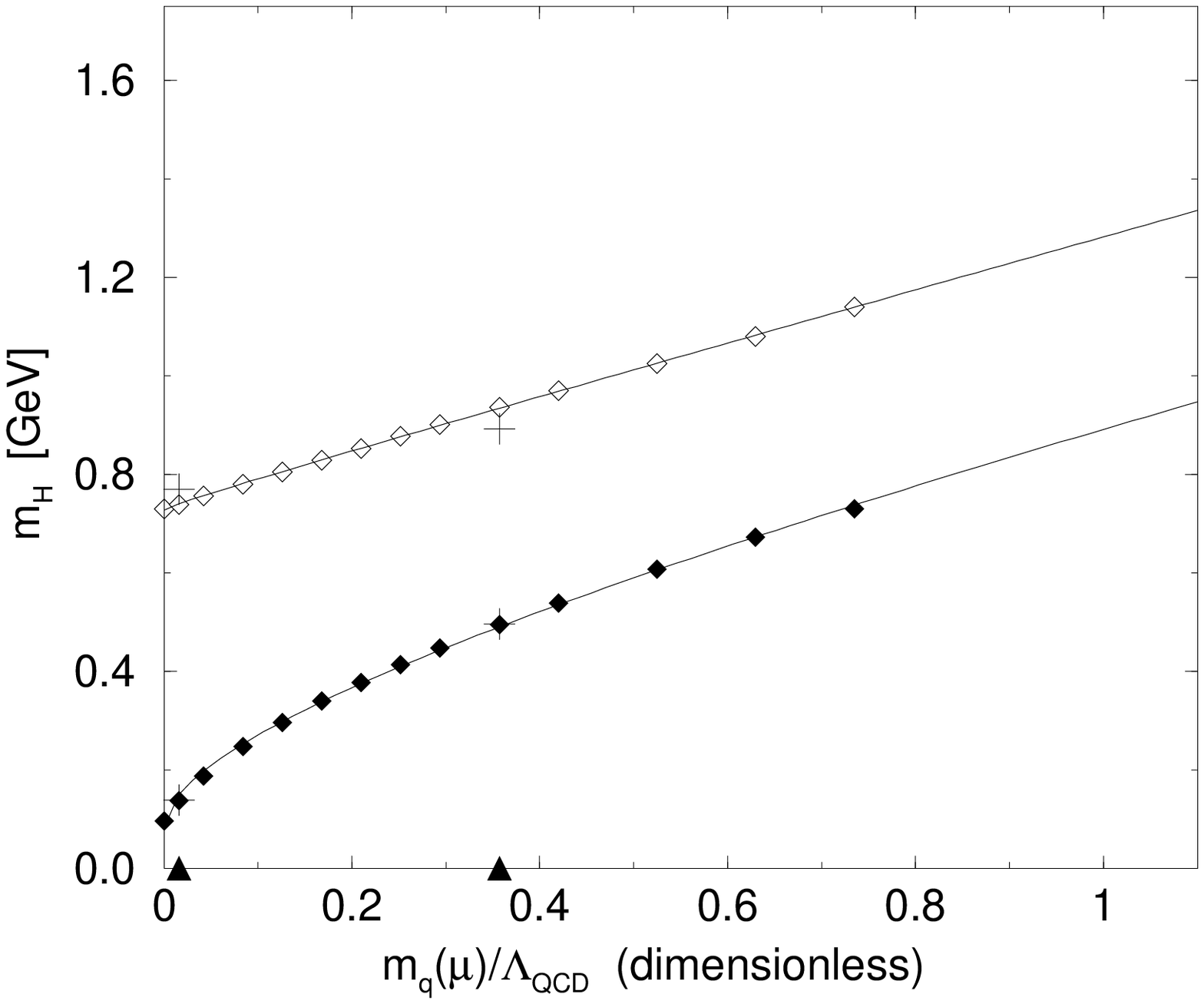,height=1.7in}
\caption{Pseudoscalar and vector meson masses as function of 
$m_q/\Lambda_{\rm QCD}$: left, mesons with equal-mass constituents;
right, mesons consisting of one $u/d$ quark and one quark with
variable mass $m_q$.  The current quark mass $m_q(\mu)$ are evaluated
at $\mu=19\;\rm{GeV}$; realistic $u/d$ and $s$ quark masses are
indicated by triangles, and the corresponding mesons by plusses.
\label{fig:psvecmss}}
\end{figure}
As can be seen from Fig.~\ref{fig:psvecmss}, the pseudoscalar meson
mass $m_{PS}$ behaves like $(m_{PS})^2 \propto m_q$ in the limit $m_q
\rightarrow 0$, as expected from the Gell-Mann--Oakes--Renner [GMOR]
relation.  Around the strange quark mass, the curve starts to deviate
significantly from this GMOR behavior, and a more linear dependence of
the meson mass on $m_q$ begins to emerge~\cite{Maris:1997cg}.  This is
in agreement with general expectations that for large $m_q$, meson
masses are simply proportional to the current quark mass.  The vector
meson masses show a much more linear dependence on $m_q$.  Over the
current mass range explored, our results for both the vector and the
pseudoscalar mesons can be fitted by a form \mbox{$m_H = \alpha +
\beta \sqrt{m_q/\Lambda_{\rm QCD}} + \gamma \, m_q/\Lambda_{\rm QCD}$}, 
with the parameters of Table~\ref{tab:mfit}.
\begin{table}[b]
\caption{Fitted parameters for the mesons mass dependence on 
$m_q$ (in GeV).
\label{tab:mfit}}
\vspace{0.2cm}
\begin{center}
\begin{tabular}{|l|llll|} \hline
                &$m_\pi$& $m_\rho$&$m_{K}$&$m_{K^*}$
\\ \hline
$\alpha$        & 0.0   & 0.703 & 0.083 & 0.728  \\
$\beta $        & 1.04  & 0.25  & 0.50  & 0.03   \\
$\gamma$        & 0.21  & 0.61  & 0.31  & 0.52   \\ \hline
\end{tabular}
\end{center}
\end{table}

The decay constants all grow with $m_q$, at least in the range of
current quark masses explored.  In the heavy quark limit, one can show
analytically that the pseudoscalar and vector decay constants vanish
like $1/\sqrt{m_q}$, as expected from heavy quark effective
theory~\cite{Ivanov:1999ms}.  The turn-over point for this behavior 
is larger than three times the strange quark mass, at least in the
present model.  The details of this turn-over do of course depend on
the details of the effective interaction.  However, the qualitative
behavior, both of the meson masses and of the decay constants, is
expected to be model-independent.

\subsection{Meson Form Factors}
\begin{figure}[t]
\psfig{figure=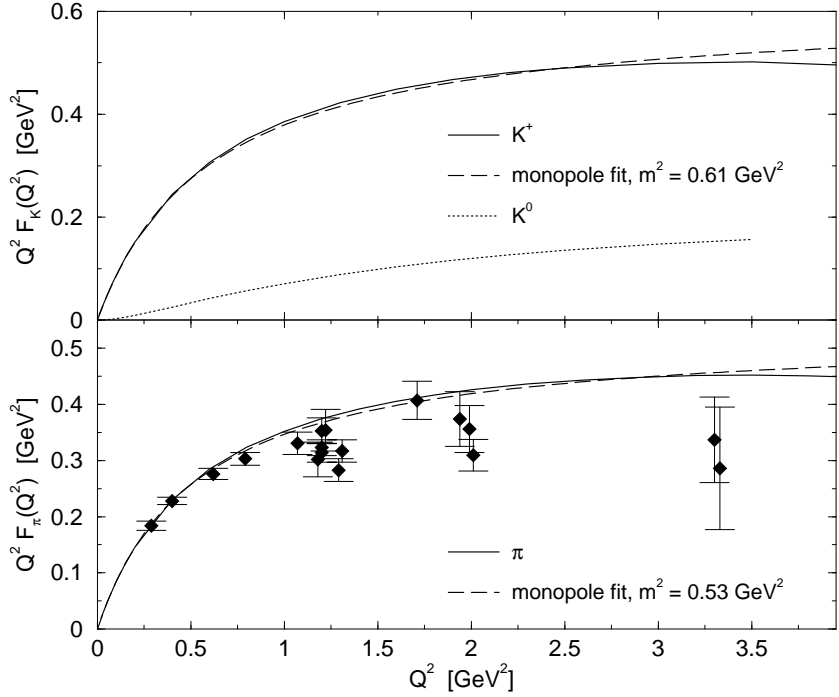,height=3.7in}
\caption{The electromagnetic form factors of the light pseudoscalar 
mesons~\protect\cite{Maris:2000sk}.  The pion data shown here are from
an analysis by Bebek {\it et al.}~\protect\cite{Bebek:1978pe}.  Our
results are in good agreement with preliminary new data from JLab for
the pion~\protect\cite{Volmer}; new results for $F_{K^+}$ are
also expected~\protect\cite{Niculescu:1998zj}.
\label{fig:Q2pikaff}}
\end{figure}

Meson electromagnetic form factors in impulse approximation are
described by two diagrams, with the photon coupled to the quark and to
the antiquark respectively.  We can define a form factor for each of
these diagrams~\cite{Maris:2000sk}, e.g.
\begin{eqnarray}
 2\,P_\nu\,F_{a\bar{b}\bar{b}}(Q^2) &=&
        N_c\int\!\!\frac{d^4q}{(2\pi)^4}
        \,{\rm Tr}\big[ S^a(q) \, \Gamma_{\rm ps}^{a\bar{b}}(q,q_+;P_-) 
\nonumber \\ &&{}\times
        S^b(q_+) \, i \Gamma^{b}_\nu(q_+,q_-)\, S^b(q_-) \,
        \bar\Gamma_{\rm ps}^{a\bar{b}}(q_-,q;-P_+) \big] \;, 
\end{eqnarray}
where \mbox{$q = k+\frac{1}{2}P$}, 
\mbox{$q_\pm = k-\frac{1}{2}P \pm \frac{1}{2}Q$},  
\mbox{$P_\pm = P \pm \frac{1}{2}Q$}.  We work in the isospin symmetry
limit, so for the pion form factor we have \mbox{$F_{\pi}(Q^2) =
F_{u\bar{u}u}(Q^2)$}.  The charged and neutral kaon form factors are 
given by \mbox{$F_{K^+}= \frac{2}{3}F_{u\bar{s}u} + 
\frac{1}{3}F_{u\bar s \bar s}$} and \mbox{$F_{K^0} = 
-\frac{1}{3}F_{d\bar{s}d} + \frac{1}{3}F_{d\bar s\bar s}$} respectively.
Our results for $Q^2 F_\pi$, $Q^2 F_{K^+}$, and $Q^2 F_{K^0}$ are shown
in Fig.~\ref{fig:Q2pikaff}, with the charge radii are given in
Table~\ref{tab:summary}.  
\begin{table}[b]
\caption{Overview of calculated meson charge radii in ${\rm fm}^2$,
with expt. data~\protect\cite{A86A86KM78}, and meson coupling constants,
with expt. values extracted from decay widths~\protect\cite{PDG}.
\label{tab:summary}}
\vspace{0.2cm}
\begin{center}
\begin{tabular}{|l|llll|llll|} \hline
& $r^2_\pi$ & $r^2_{K^+}$ & $r^2_{K^0}$ & $r^2_{\pi\gamma\gamma}$ 
& $g_{\pi\gamma\gamma}$& $g_{\rho\pi\gamma}$&
                        $g_{\rho\pi\pi}$& $g_{\phi K K}$ \\ \hline
expt. data      & 0.44 & 0.34 & -0.054 & 0.42 & 0.50 & 0.57 & 6.02 & 4.64 \\ 
calculation     & 0.45 & 0.38 & -0.086 & 0.39 & 0.50 & 0.54 & 4.85 & 4.63 \\
\hline
\end{tabular}
\end{center}
\end{table}

Up to about $Q^2 = 3\,{\rm GeV}^2$, our results for $F_\pi$ and $F_K$
can be fitted quite well by a monopole.  Asymptotically, these
functions behave like \mbox{$Q^2 F(Q^2) \rightarrow c$} up to
logarithmic corrections.  However, numerical limitations prevent us
from accurately determining these constants. Around $Q^2 = 3\,{\rm
GeV}^2$, our result for $Q^2 F_\pi$ is well above the pQCD
result~\cite{Farrar:1979aw} \mbox{$16 \pi f_\pi^2 \alpha_s(Q^2) \sim
0.2 \,{\rm GeV}^2$}, and clearly not yet asymptotic.  If we continue
the calculation for the electromagnetic form factor into the timelike
region, we find a pole at the mass of the vector meson bound states.
Using the behavior of $F_{u\bar{u}u}$ and $F_{u\bar{s}\bar{s}}$ around
this pole, we can extract the coupling constants $g_{\rho\pi\pi}$ and
$g_{\phi K K}$ respectively~\cite{MTinprogress}, which govern the
strong decays $\rho\rightarrow \pi\pi$ and $\phi\rightarrow K K$.  The
results from this analysis are also given in Table~\ref{tab:summary},
and are reasonably close to the experimental data.

\begin{figure}[t]
\psfig{figure=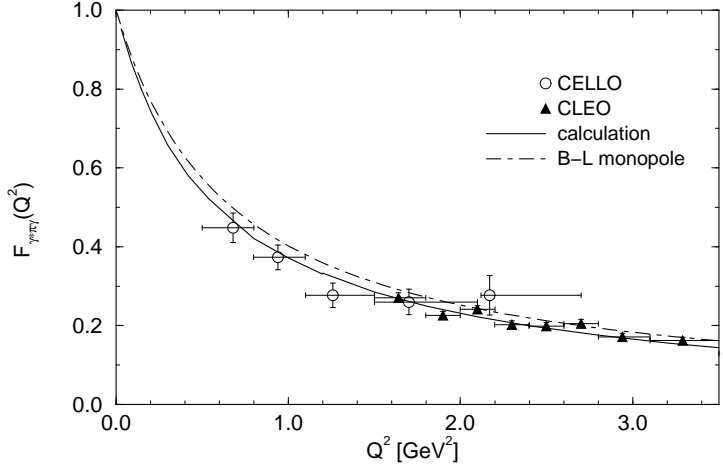,height=2.5in}
\caption{The $\gamma^\star\,\pi\gamma$ form 
factor~\protect\cite{Maris:2000wz}, with data from CLEO and
CELLO~\protect\cite{cellocleo}.  \label{fig:piggff}}
\end{figure}

The impulse approximation for the $\gamma^\star\pi\gamma$ vertex 
with $\gamma^\star$ momentum $Q$ is
\begin{eqnarray}
\lefteqn{ \Lambda_{\mu\nu}(P,Q)=i\frac{\alpha }{\pi f_{\pi }}
        \,\epsilon_{\mu \nu \alpha \beta }\,P_{\alpha }Q_{\beta }
        \, g_{\pi\gamma\gamma}\,F_{\gamma^\star\pi\gamma}(Q^2)  } \\
& & \nonumber
        =\frac{N_c}{3}\, \int\!\frac{d^4q}{(2\pi)^4}
        {\rm Tr}\left[S(q)\, i\Gamma_\nu (q,q')\,S(q')\, 
                i\Gamma_\mu (q',q'')\,S(q'')\,\Gamma_\pi(q'',q;P)\right] \;.
\end{eqnarray}
where the momenta follow from momentum conservation.  In the chiral
limit, the value at $Q^2 = 0$, corresponding to the decay \mbox{$\pi^0
\rightarrow \gamma\gamma$}, is given by the axial anomaly and its value
\mbox{$g^{0}_{\pi\gamma\gamma}=\frac{1}{2}$} is a direct consequence of 
only gauge invariance and chiral symmetry; this value is reproduced by
our calculations~\cite{Maris:1998hc} and corresponds well with the
experimental width of $7.7~{\rm eV}$.  In Fig.~\ref{fig:piggff} we show
our results with realistic quark masses, normalized to the experimental
$g_{\pi\gamma\gamma}$.
The large-$Q^2$ behavior of $F_{\gamma^\star\pi\gamma}$ lies in
between monopoles fitted to the Brodsky--Lepage pQCD asymptotic
limit~\cite{Lepage:1980fj}, $8 \pi^2 f_\pi^2$, and the limit obtained
from a naive analysis of the impulse approximation within the DSE
approach~\cite{Kekez:1999rw,dsegpglimit}, $\frac{16}{3}\pi^2 f_\pi^2$.
Our results for $F_{\gamma^\star\pi\gamma}$ are compatible with an
estimate for the asymptotic behavior by Anikin {\it et
al.}~\cite{Anikin:2000cx}.

\section{Finite Temperature Extension}
\label{finiteT}
In the Matsubara formalism, the quark propagator can be written as
\begin{eqnarray}
S^{-1}(\vec{p},\omega_k)  & = & 
	i\vec{\gamma}\cdot \vec{p} \,A(p^2,\omega_k^2)
        + i\gamma_4\,\omega_k \,C(p^2,\omega_k^2) + B(p^2,\omega_k^2) 
\\
        & = & Z_2^A \,i\vec{\gamma}\cdot \vec{p} 
        + Z_2^C \, i\gamma_4\,\omega_k + Z_4 m(\mu) 
	+ \Sigma(\vec{p},\omega_k)\,,
\label{qDSE} 
\end{eqnarray}
where $\omega_k= (2 k + 1)\,\pi T$ is the fermion Matsubara frequency
and $T$ the temperature.  The functions $A$ and $C$ are renormalized
such that $A(p^2,\omega_0^2) = 1 = C(p^2,\omega_0^2)$, and
$B(p^2,\omega_0^2) = m(\mu)$ at $p^2 + \omega_0^2 = \mu^2$.  The
approximations we use are similar to the $T=0$ studies discussed in
the previous sections.  The rainbow-ladder truncation for the
self-energy is
\begin{equation}
\label{sigmap}
 \Sigma(\vec{p},\omega_k) = \frac{4}{3} T \sum_{l=-\infty}^\infty 
        \int^{\Lambda}\frac{d^3q}{(2\pi)^3} 
        g^2\,D_{\mu\nu}(\vec{p}-\vec{q},\Omega_{k-l}) \gamma_\mu
        S(\vec{q},\omega_l)\gamma_\nu\,,
\label{regself}
\end{equation}
where $g^2 D_{\mu\nu}$, the effective interaction or effective gluon
propagator, at finite T is taken to be the minimal extension of a
$T=0$ model~\cite{Maris:2000ig,Bender:1996bm,Holl:1999qs} with all
parameters fixed at $T=0$.  This truncation leads to a second-order
chiral phase transition
\begin{equation}
 \langle \bar q q \rangle^0 
        \propto (1-T/T_c)^{\beta}\,,\;\; T/T_c\uparrow 1\,,
\end{equation}
where $\beta$ is the critical exponent for chiral symmetry
restoration.  Independent of the details of the effective interaction,
we find mean field behavior, $\beta = 1/2$, in rainbow-ladder
models~\cite{Holl:1999qs}.  However, one has to go to extremely small
current quark masses, several orders of magnitude smaller than
realistic $u/d$ quark masses, to observe this mean-field behavior.
The critical temperature is sensitive to details of the model, but
with the model parameters fitted to $T=0$ pion observables, one
typically finds $T_c \sim 0.15$~GeV.

\subsection{Meson correlations at finite temperature}
In order to study meson correlations~\cite{Maris:2000ig} we solve the
inhomogeneous scalar and pseudoscalar vertices $\Gamma_{S,PS}$ for the
lowest external (meson) Matsubara mode, $\Omega_0 = 0$.  Bound states
will appear as poles in these vertices; for this choice of the
external momentum these poles correspond to the spatial modes, and
thus the masses to the spatial
masses~\cite{Maris:2000ig,Blaschke:2000gd}, which are the type of
masses that are usually calculated in lattice simulations.

For $\hat m =0$ the $T=0$ inhomogeneous BSEs exhibit poles at $ m_\pi
= 0$ and $m_\sigma= 0.56\,{\rm GeV}$, with $m_\sigma = 0.59\,$GeV at
$\hat m = 5.7\,$MeV.  This low-mass scalar is typical of the
rainbow-ladder truncation, although there is some model sensitivity.
The rainbow-ladder truncation yields degenerate isoscalar and
isovector bound states, and ideal flavor mixing in the $3$-flavor
case; improvements beyond the ladder truncation are required in order
to describe the observed scalar mesons.  Furthermore, in the
isoscalar-scalar channel, it will be necessary to include couplings to
the dominant $\pi\pi$ mode~\cite{Pennington:2000hj} because of the
large $\sigma$ width.  In the absence of such corrections, the
$\sigma$ properties elucidated herein are strictly only those of an
idealized chiral partner of the $\pi$.

\begin{figure}[t]
\psfig{figure=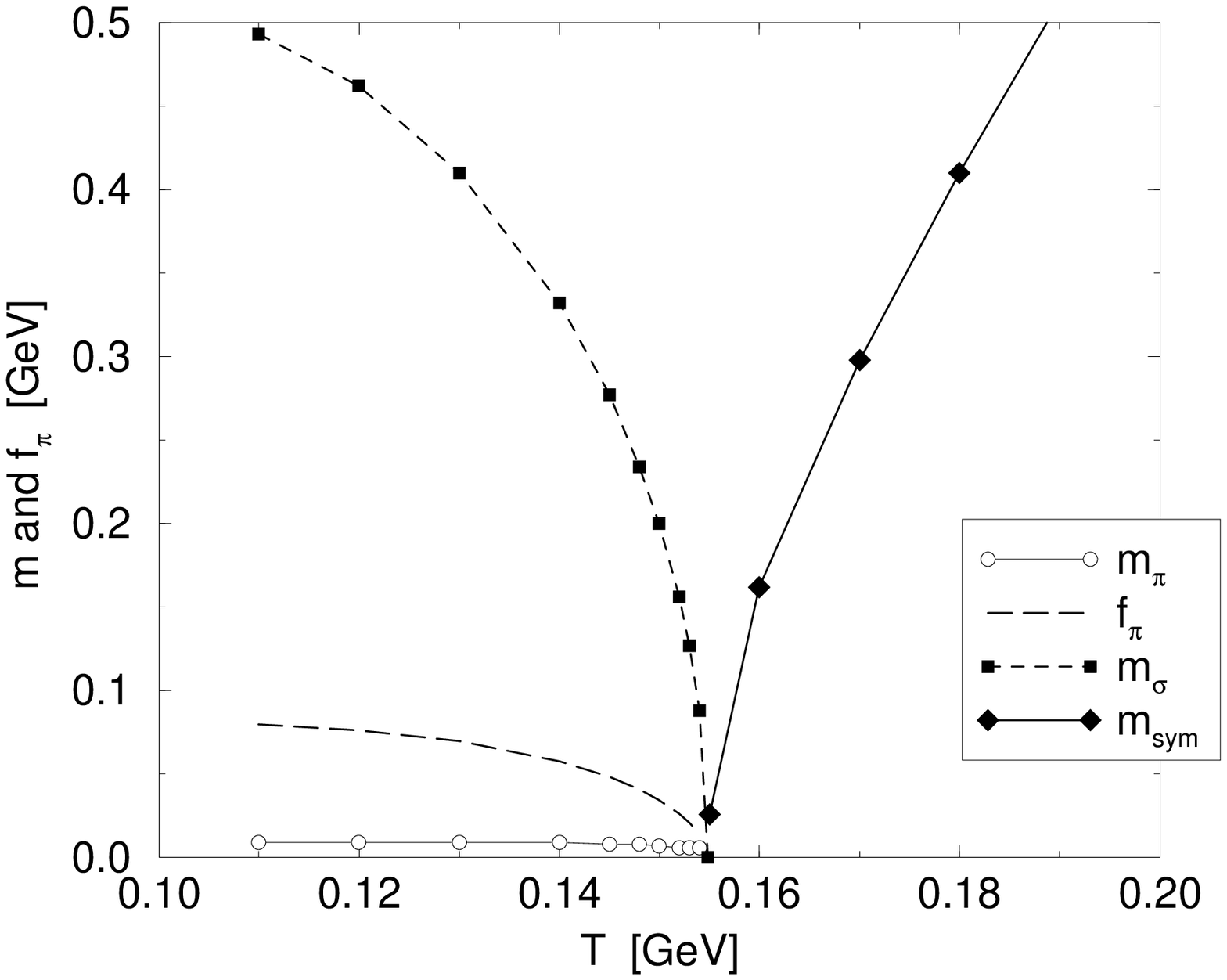,height=1.7in}
\vspace*{-1.7in}

\hspace*{2.4in}\psfig{figure=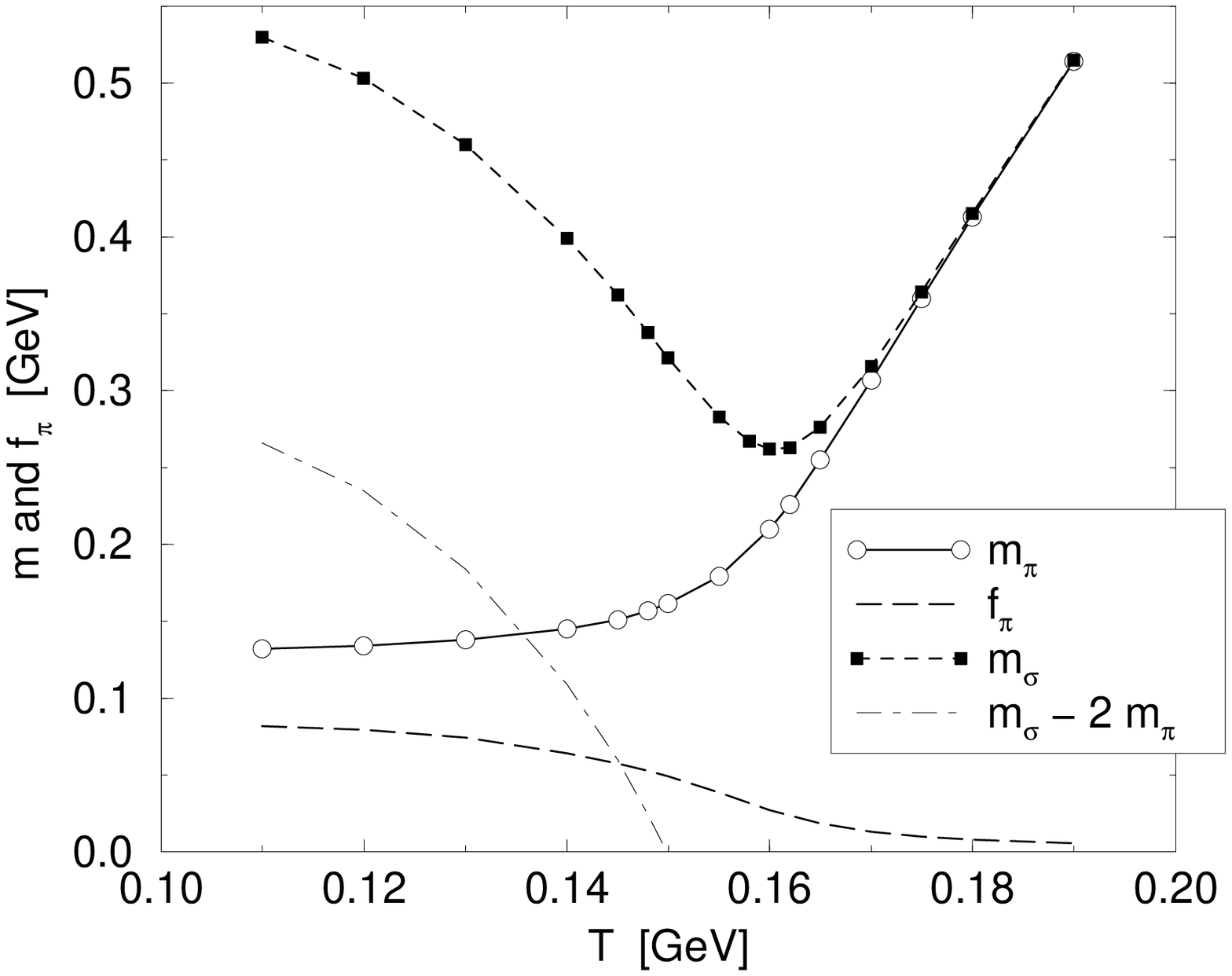,height=1.7in}
\caption{The $\pi$ and $\sigma$ mass, decay constant as function of $T$,
both in the chiral limit (left) and with realistic current quark mass
(right)~\protect\cite{Maris:2000ig}.  \label{fig:pisigT}}
\end{figure}

These poles evolve with $T$ in such a way~\cite{Maris:2000ig} that at
the critical temperature, we have degenerate, massless pseudoscalar
and scalar bound states, see Fig.~\ref{fig:pisigT}.  Above $T_c$,
these bound states persist, becoming increasingly massive with
increasing $T$ and approaching $2\pi T$, the thermal mass of two free
quarks.  Similar features are also observed in numerical simulations
of lattice-QCD~\cite{echaya2}.  Above $T_c$ all but the leading Dirac
amplitude vanishes and the surviving pseudoscalar amplitude is
pointwise identical to the surviving scalar one.  These results
indicate that the chiral partners are {\it locally} identical above
$T_c$, they do not just have the same mass.  Approaching $T_c$ from
below, our analysis yields for the $\sigma$ mass $ m_\sigma
\propto (1-T/T_c)^{\beta}$ within numerical errors.  Note that this 
is the same behavior as the chiral condensate and other, equivalent
order parameters for chiral symmetry restoration, such as the decay
constant $f_\pi$.

For nonzero current-quark masses chiral symmetry restoration is
exhibited as a crossover rather than a phase transition.  That the
transition has become a crossover is evident in the behavior of
$f_\pi$, see the right panel of Fig.~\ref{fig:pisigT}.  The $\sigma$
mass exhibits a dip around the phase transition.  The scalar and
pseudoscalar meson masses become indistinguishable at $T \sim 1.2
\,T_c$.

With the BSAs obtained from the homogeneous BSE, we can now calculate
the $T$-dependence of other meson properties~\cite{Maris:2000ig}.  The
$T$-dependence of $g_{\pi\gamma\gamma}$ is calculated in a similar
fashion as at $T=0$ in the previous section.  At $T=0$ this coupling
saturates the Abelian anomaly: $g_{\pi\gamma\gamma}= \frac{1}{2}$.
The interesting quantity is: $g^0_{\pi\gamma\gamma}/f_\pi^0$, which
vanishes at $T_c$ in the chiral limit~\cite{pisarski}:
$g^0_{\pi\gamma\gamma}$ goes to zero linearly.  Thus, in the chiral
limit, the coupling to the dominant decay channel closes for both
charged and neutral pions.  For $\hat m
\neq 0$ both the coupling: $g_{\pi\gamma\gamma}/f_\pi$, and the 
corresponding width exhibit a crossover with a slight enhancement in
the width as $T\to T_c$ due to the increase in $m_\pi$.

Also the isoscalar-scalar-$\pi\pi$ strong coupling vanishes at $T_c$
in the chiral limit, which can be traced to
$B_{\hbox{\scriptsize{chiral}}} \to 0$.  For $\hat m \neq 0$, the
coupling reflects the crossover.  However, that is irrelevant because
the width vanishes just below $T_c$ where the $\sigma$ meson mass
falls below $2 m_\pi$ and the phase space factor vanishes.  A similar
phenomenon occurs for the $\rho\pi\pi$ decay width: with a separable
model for the effective interaction it was
shown~\cite{Blaschke:2000gd} that slightly above $T_c$ the spatial
mass of the $\rho$ becomes smaller than $2 m_\pi$, and thus the strong
decay of a $\rho$ into pions becomes impossible because of phase space
consideration.  However, one should keep in mind that these are the
spatial masses, and that effects such as collisional broadening are
not yet taken into account.

\section*{Acknowledgments}
I would like to thank Wolfgang Lucha and Kim Maung Maung for their
efforts in organizing this conference.  I am also grateful to the
Erwin Schroedinger Institute for Mathematical Physics in Vienna for
their hospitality and financial support.  Most of the work discussed
here was done in collaboration with Peter Tandy and Craig Roberts; I
would also like to thank Tony Williams, Sebastian Schmidt, Dennis
Jarecke, and David Blaschke for useful discussions.  This work was
funded by the National Science Foundation under grants
Nos. INT-9603385 and PHY97-22429, and benefited from the resources of
the National Energy Research Scientific Computing Center.

\section*{References}

\end{document}